\documentclass[twocolumn, prl, showpacs,superscriptaddress]{revtex4}
\usepackage{graphicx}
\usepackage{amsmath}
\usepackage{bm}
\usepackage{float}
\usepackage{color}
\usepackage{sidecap}

\begin{document}
\title{Dynamically generating arbitrary spin-orbit couplings for neutral atoms}
\author{Z. F. Xu}
\author{L. You}
\affiliation{State Key Laboratory of Low Dimensional Quantum Physics,
Department of Physics, Tsinghua University, Beijing 100084, China}

\date{\today}

\begin{abstract}
Spin-orbit coupling (SOC) can give rise to interesting physics, from
spin Hall to topological insulators, normally in condensed matter
systems. Recently, this topical area has extended into atomic
quantum gases in searching for artificial/synthetic gauge
potentials. The prospects of tunable interaction and quantum state
control promote neutral atoms as nature's quantum emulators for SOC.
Y.-J. Lin {\it et al.} recently demonstrated a special form of the
SOC $k_x\sigma_y$: which they interpret as an equal superposition of
Rashba and Dresselhaus couplings, in bose condensed atoms [Nature
(London) \textbf{471}, 83 (2011)]. This work reports an idea capable
of implementing arbitrary forms of SOC by switching between two
pairs of Raman laser pulses like that used by Lin {\it et al.}.
While one pair affects $k_x\sigma_y$ for some time, a second pair
creates $k_y\sigma_y$ over other times with Raman pulses from
different directions and a subsequent spin rotation into $\pm
k_y\sigma_x$. With sufficient many pulses, the effective actions
from different durations are small and accumulate in the same
exponent despite that $k_x\sigma_y$ and $\pm k_y\sigma_x$ do not
commute. Our scheme involves no added complication, and can be
demonstrated within current experiments. It applies equally to
bosonic or fermionic atoms.
\end{abstract}


\maketitle

{\it Introduction.} Atomic quantum gases are increasingly viewed as
favored model systems for emulating condensed matter physics.
Optical lattices resulting from ac Stack shifts to atomic levels,
are easily implemented with coherent laser beams, which confine
atoms like electrons in solid states. An interesting topic concerns
strong correlations as in integer/fractional quantum Hall effect and
the analogous spin Hall effect. The standard description for the
former involves U(1) Abelian gauge fields, which can be simulated in
neutral atoms through rotation \cite{fetter2009,cooper2008} or
adiabatic translations in far-off-resonant laser fields
\cite{dalibard2010,juzeliunas2005,gunter2009,lin2009}. Non-Abelian
gauge fields, {\it e.g.}, as in spin-orbit coupling (SOC)
\cite{ruseckas2005,juzeliunas2010,lin2011,sau2011,campbell2011},
enable richer possibilities like fractional quantum Hall states. As
a result, active researches are targeting the implementations of
(SOC) in simple atomic systems.

For atoms with multiple internal states, or (pseudo-spin) spinor
degrees of freedom, SOC changes single particle spectra and competes
with density-density or spin-dependent interactions, (i.e.,
spin-exchange and singlet-pairing interactions). Strong correlations
often lead to exotic ground states
\cite{stanescu2008,wang2010,ho2010,yip2011,xu2011,kawakami2011,wu2011,zhang2011,hu2011,sinha2011},
such as the plane-wave phase and the striped phase discovered
recently in pseudo spin-1/2 \cite{wang2010,ho2010,yip2011} or spin-1
condensates \cite{wang2010}. Other examples offer the triangular-latticed
phase or square-latticed phase in spin-2 condensates with
axisymmetric SOC \cite{xu2011,kawakami2011}. In a recent experiment,
the JQI group of Spielman observed both Abelian \cite{lin2009} and
non-Abelian \cite{lin2011} gauge fields in a pseudo spin-1/2 atomic
Bose gas, albeit in a special form $\propto k_x\sigma_y$ of SOC,
which is an equally weighted sum of Rashba ($\propto k_x\sigma_y-k_y\sigma_x$)
and Dresselhaus ($\propto k_x\sigma_y+k_y\sigma_x$) couplings
\cite{lin2011}. More generally, a SOC form of continuous rotation
symmetry, or an arbitrary weighted sum of Rashba and Dresselhaus
couplings, exists in solid-state materials.

Several existing theoretical proposals are capable of implementing
SOC with rotation symmetry in laser atom coupled models. For
instance, in a tripod scheme \cite{ruseckas2005}, when one-photon
resonant couplings between the three lower-energy states and a higher-energy one
are allowed, two dark states emerge, although spontaneous emission
is always a cause of concern in this case. D. L. Campbell {\it et
al.} \cite{campbell2011} proposed an alternative scheme by
cyclically coupling three or four ground or metastable internal
states. With sufficient laser intensities, the above
induced SOCs can possess a continuous rotation symmetry.
Another scheme by Jay D. Sau {\it et al.} \cite{sau2011}
employs an effective two-dimensional periodic potential created from
two laser beams and their reflected lights propagating along
$\hat{x}$ and $\hat{y}$ directions in $^{40}$K atoms. In the limit
of small Raman coupling, their corresponding effective SOC
is of a pure Rashba type in the first Brillouin zone.

In this Letter, we describe a dynamic approach for implementing
rotational symmetric SOC of arbitrary forms within a pseudo-spin 1/2 atomic system.
We adopt the JQI model and start from the simple SOC they proposed
and recently demonstrated \cite{lin2011}.
The key to our idea is optimal control theory applied with repeated laser
pulses to rotate atomic pseudo-spins. Our idea works for both atomic fermions
and bosons, and can be easily adopted to other atomic models.
Thus it constitutes a powerful new direction for engineering synthetic
atomic gauge potentials.

The equally weighted sum of Rashba and Dresselhaus types SOC of
$k_x\sigma_y$ \cite{lin2011}, can be rotated into a form $\propto\pm k_y\sigma_x$,
by performing single atom spin rotation through a Rabi pulse.
Such a coherent control idea when repeated over time,
can  realize $k_x\sigma_y$
and $\pm k_y\sigma_x$ types SOC in subsequent time intervals of duration $\delta t$.
The resulting dynamics is then described respectively by an effective Hamiltonian
with pure Rashba or Dresselhaus SOC with the first order approximation for small $\delta t$.
The accompanied change of atomic momentum, can be nullified through a variety
of means as we describe below step by step.
We start with a review of the experiment by Y.-J. Lin {\it et al.} \cite{lin2011},
which helps to introduce our idea.

{\it The JQI protocol.}
Consider a $F=1$ atomic Bose-Einstein condensate (BEC) under
a bias magnetic field along $\hat{z}$ located
at the intersection of two
Raman laser beams propagating along $\hat{y}+\hat{z}$ and $-\hat{y}+\hat{z}$,
with angular frequencies $\omega_L$ and $\omega_{L}+\Delta\omega_{L}$, respectively.
The two laser beams affect two photon resonant Raman coupling ($\Omega_R$) between
nearby ground Zeeman states, far detuned from the excited states.
Effectively, such a coupling scheme produces an artificial magnetic field
along the $x$-axis direction with the resulting Hamiltonian
$\hat{H}_R=\Omega_RF_x\cos(2k_L\hat{y}+\Delta\omega_Lt)$,
where $F_{x,y,z}$ are $3\times3$ spin-1 matrices, $k_L=\sqrt{2}\pi/\lambda$ with
$\lambda$ is the laser wavelength, and $E_L=\hbar^2k_L^2/2m$, the unit of
photon recoil energy.
In explicit forms, after adiabatically eliminating excited states,
the total Hamiltonian becomes
\begin{widetext}
\begin{eqnarray}
  \hat{H}_3=\frac{\hbar^2\hat{\mathbf{k}}^2}{2m}+
  \left(
  \begin{array}{ccc}
    E_+ & 0 & 0 \\
    0 & E_0 & 0 \\
    0 & 0 & E_{-} \\
  \end{array}
  \right)
  +\frac{\Omega_R}{\sqrt{2}}
  \left(
  \begin{array}{ccc}
    0 & \cos(2k_L\hat{y}+\Delta\omega_Lt) & 0 \\
    \cos(2k_L\hat{y}+\Delta\omega_Lt) & 0 & \cos(2k_L\hat{y}+\Delta\omega_Lt) \\
    0 & \cos(2k_L\hat{y}+\Delta\omega_Lt) & 0 \\
  \end{array}
  \right),
  \label{ham3}
\end{eqnarray}
\end{widetext}
where $E_+$, $E_0$ and $E_-$ are Zeeman (eigen-) energies of $M_F=1,0,-1$ spin states,
respectively. Under the rotating wave approximation, it turns into
\begin{eqnarray}
  \hat{H}_3&=&\frac{\hbar^2\hat{\mathbf{k}}^2}{2m}+
  \left(
  \begin{array}{ccc}
    E_+ & 0 & 0 \\
    0 & E_0 & 0 \\
    0 & 0 & E_{-} \\
  \end{array}
  \right)
  \nonumber\\
  &&+\frac{\Omega_R}{2}F_x\cos(2k_L\hat{y}+\Delta\omega_Lt)\nonumber\\
  &&-\frac{\Omega_R}{2}F_y\sin(2k_L\hat{y}+\Delta\omega_Lt).
  \label{ham32}
\end{eqnarray}
Further introduce a frame transformation: $\tilde{\psi}=e^{-iF_z
\Delta\omega_Lt}\psi$, where $\psi$ and $\tilde{\psi}$ are the wave
functions in the laboratory and transformed frames, respectively, we
arrive at the Hamiltonian
\begin{eqnarray}
  \hat{H}_3&=&\frac{\hbar^2\hat{\mathbf{k}}^2}{2m}+
  \left(
  \begin{array}{ccc}
    2\hbar\omega_q+3\delta/2 & 0 & 0 \\
    0 & \delta/2 & 0 \\
    0 & 0 & -\delta/2 \\
  \end{array}
  \right)+E_0-\delta/2
  \nonumber\\
  &+&\frac{\Omega_R}{2}F_x\cos(2k_L\hat{y})-\frac{\Omega_R}{2}F_y\sin(2k_L\hat{y}),
  \label{ham33}
\end{eqnarray}
where $\hbar\omega_Z=E_--E_0$, $\hbar\Delta\omega_L=\hbar\omega_Z+\delta$,
$E_0-E_+=\hbar\omega_Z-2\hbar\omega_q$, $\delta$ is detuning and $\hbar\omega_q$ is the quadratic
Zeeman shift. When $\hbar\omega_q$ is sufficiently large and the Raman coupling
$\Omega=\Omega_R/\sqrt{2}$ is small,
we neglect the state $|M_F=1\rangle$ and a constant term $E_0-\delta/2$.
The effective Hamiltonian for the remaining two nearly degenerate states becomes
\begin{eqnarray}
  \hat{H}_2&=&\frac{\hbar\mathbf{k}^2}{2m}+\frac{\delta}{2}\sigma_z
  +\frac{\Omega}{2}\sigma_x\cos(2k_L\hat{y})-\frac{\Omega}{2}\sigma_y\sin(2k_L\hat{y})
  \nonumber\\
  &=&e^{ik_L\hat{y}\sigma_z}\left(\frac{\hbar^2\mathbf{k}^2}{2m}
  +\frac{\delta}{2}\sigma_z+\frac{\Omega}{2}\sigma_x
  +2\frac{\hbar^2k_L\hat{k}_y}{2m}\sigma_z+E_L
  \right)\nonumber\\
  &&\times e^{-ik_L\hat{y}\sigma_z},
  \label{ham2}
\end{eqnarray}
where the second line shows an explicit SOC term $\propto \hat{k}_y\sigma_z$
when viewed after a unitary transformation.

\begin{figure}[H]
\centering
\includegraphics[width=2.5in]{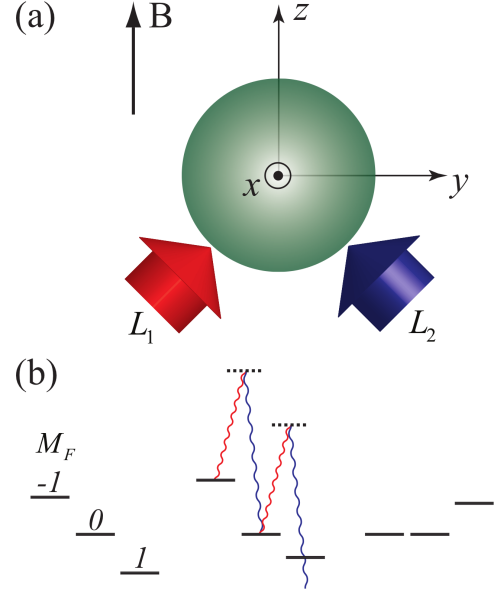}
\caption{(Color online). (a) A schematic illustration of
the JQI implementation for SOC \cite{lin2011}, where a
$F=1$ atomic Bose-Einstein condensate interacts with a bias magnetic
field along $\hat{z}$ and two Raman laser beams propagating along
$\hat{y}+\hat{z}$ and $-\hat{y}+\hat{z}$, with angular frequencies
$\omega_L$ and $\omega_{L}+\Delta\omega_{L}$, respectively. (b)
LEFT: Linear Zeeman shifts of the three hyperfine spin states. MIDDLE:
Zeeman shifts of the three hyperfine spin states including both linear and
quadratic terms. RIGHT: Zeeman shifts in the rotating frame (with
frequency $\Delta\omega_L F_z$) of the pseudo-spin pointing along
$\hat{z}$ .
 }
\label{fig1}
\end{figure}

{\it Dynamically generating arbitrary SOC.} Our protocol for
implementing the Rashba type SOC is illustrated below in Fig. \ref{fig2}.
It relies on our ability of being able to switch atomic pseudo-spin from
along $z$- to along $y$-axis (and {\it vice versa})
using Raman pulses. In the
first half period, Raman lasers $L_1$ and $L_2$ are turned on.
In the second half, $L_3$ and $L_2$ are turned on instead. $L_3$ is
the same as $L_1$ except it propagates along opposite direction.
At the middle point, we pulse on an extra $\pi/2$ pulse to
rotate the pseudo-spin from $y$- to $z$-axis, described
by the operator $\exp[-i(\sigma_x/2)\pi/2]$ in the transformed
frame, or the operator
$\exp[iF_z\Delta\omega_Lt]\exp[-i(\sigma_x/2)\pi/2]\exp[-iF_z
\Delta\omega_Lt]$ in the lab frame; in the end of each
period, we pulse on an $-\pi/2$ pulse for the reverse rotation. Both
spin rotation pulses can be accomplished with either Raman coupling
from appropriately detuned lasers or rf plus microwave coupling
between the two remaining internal states.

In the first half, the system is
then governed by $\hat{H}_3$ of the Eq.~(\ref{ham33})
as in the JQI experiment \cite{lin2011}.
In the second half, the Hamiltonian
in the transformed frame following that in the Eq.~(\ref{ham33}) becomes
\begin{eqnarray}
  \hat{H}'_3&=&\frac{\hbar^2\hat{\mathbf{k}}^2}{2m}+
  \left(
  \begin{array}{ccc}
    2\hbar\omega_q+3\delta/2 & 0 & 0 \\
    0 & \delta/2 & 0 \\
    0 & 0 & -\delta/2 \\
  \end{array}
  \right)+E_0-\delta/2
  \nonumber\\
  &&+\frac{\Omega_R}{2}F_x\cos(2k_L\hat{z})+\frac{\Omega_R}{2}F_y\sin(2k_L\hat{z}),
  \label{ham34}
\end{eqnarray}
which completes one period of our prescribed protocol.
For large $\hbar\omega_q$ and small $\Omega$, the same condition
as in Ref. \cite{lin2011}, the effective Hamiltonian for the
reduced two-state model becomes
\begin{eqnarray}
  \hat{H}'_2&=&\frac{\hbar\mathbf{k}^2}{2m}+\frac{\delta}{2}\sigma_z
  +\frac{\Omega}{2}\sigma_x\cos(2k_L\hat{z})+\frac{\Omega}{2}\sigma_y\sin(2k_L\hat{z})
  \nonumber\\
  &=&e^{-ik_L\hat{z}\sigma_z}\left(\frac{\hbar^2\mathbf{k}^2}{2m}
  +\frac{\delta}{2}\sigma_z+\frac{\Omega}{2}\sigma_x
  -2\frac{\hbar^2k_L\hat{k}_z}{2m}\sigma_z+E_L
  \right)
  \nonumber\\
  &&\times e^{ik_L\hat{z}\sigma_z}.
  \label{ham22}
\end{eqnarray}
The pair of $\pi/2$ pulse (before) and $-\pi/2$ pulse (after)
affects a unitary transformation
\begin{eqnarray}
  &&e^{i(\sigma_x/2)\pi/2}\hat{H}'_2e^{-i(\sigma_x/2)\pi/2}
  \nonumber\\
  &=&\frac{\hbar\mathbf{k}^2}{2m}+\frac{\delta}{2}\sigma_y
  +\frac{\Omega}{2}\sigma_x\cos(2k_L\hat{z})+\frac{\Omega}{2}\sigma_z\sin(2k_L\hat{z})
  \nonumber\\
  &=&e^{ik_L\hat{z}\sigma_y}\left(\frac{\hbar^2\mathbf{k}^2}{2m}
  +\frac{\delta}{2}\sigma_y+\frac{\Omega}{2}\sigma_x
  +2\frac{\hbar^2k_L\hat{k}_z}{2m}\sigma_y+E_L
  \right)
  \nonumber\\
  &&\times e^{-ik_L\hat{z}\sigma_y}.
  \label{ham23}
\end{eqnarray}

\begin{figure}[tbp]
\centering
\includegraphics[width=2.6in]{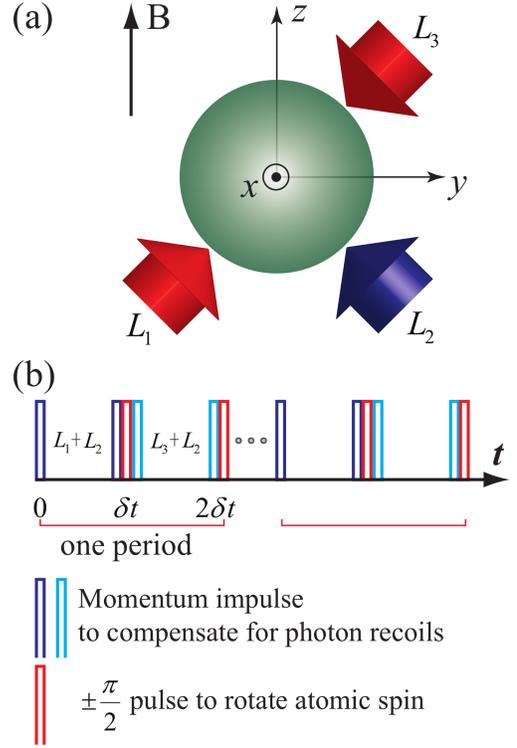}
\caption{(Color online). (a) A schematic diagram for dynamically
generating arbitrary spin-orbit coupling. (b) The pulse sequence used
to implement Rashba type SOC. The blue and cyan ones are suitable momentum
impulses used to compensate for the unwanted photon recoils, which
can be accomplished with artificial or real inhomogeneous magnetic
fields or suitably arranged state dependent Bragg pulses. The red
$\pm \pi/2$ pulse rotates atomic spin.}
\label{fig2}
\end{figure}

In suitably transformed frames, respectively with
$U_1=e^{-ik_L\hat{y}\sigma_z}$ and $U_2=e^{ik_L\hat{z}\sigma_y}$,
Eqs (\ref{ham2}) and (\ref{ham23}) reveal explicit SOC terms
$\hat{k}_y\sigma_z$ and $\hat{k}_z\sigma_y$. They cannot, however,
be simply added together in the forms above. To combine the above two
halves into a single Rashba or Dresselhaus type SOC, we have to
eliminate these unitary transformations. Both $U_1$ and $U_2$
corresponds to spin dependent phase shifts, they can be viewed as
from the impulse of an artificial or real small magnetic field along
a suitable direction and with a spatial gradient. Thus they can be
nullified by real magnetic field gradients or synthetic magnetic
field gradients generated from spatial dependent ac Stark shifts.
For instance, $U_1$ is compensated for by a magnetic field
pointing along $z$-axis and a spatial gradient ($B'$) along
$y$-axis, with an adjustable impulse over $\delta t'$ where
$E'\equiv -\mu B'$ is the appropriate Zeeman energy gradient. After
the control pulse $\delta t$, the sign of $B'$ is changed to affect
a second impulse, which then leads to the following
\begin{eqnarray}
  &&e^{-iE'\hat{y}F_z\delta t'/\hbar}e^{-i\hat{H}_3\delta t/\hbar}e^{iE'\hat{y}F_z\delta t'/\hbar}
  \nonumber\\
  &=&\exp\left\{-i\left(\frac{\hbar^2\mathbf{k}^2}{2m}+(\hbar\omega_q+\delta)F_z
  +\hbar\omega_q F_z^2+E_0\right.\right.
  \nonumber\\
  &&\qquad\left.\left.+\frac{\Omega_R}{2}F_x+2\frac{\hbar^22k_L\hat{k}_y}{2m}F_z+E_L\right)
  \delta t/\hbar\right\},
\end{eqnarray}
provided $E'\delta t'=2\hbar k_L$, where we assume $E'$ is strong
enough so that we can neglect the contribution from $\hat{H}_3$
during the short pulse $\delta t'$ ($\ll \delta t$). The effective
two-state dynamics is then approximately govern by
\begin{eqnarray}
  \exp\left\{-i\left(\frac{\hbar^2\mathbf{k}^2}{2m}+\frac{\delta}{2}\sigma_z
 +\frac{\Omega}{2}\sigma_x+2\frac{\hbar^2k_L\hat{k}_y}{2m}\sigma_z\right)\delta t/\hbar\right\}, \
\end{eqnarray}
apart from a overall phase term involving a constant energy in the exponent.
Similarly, $U_2$ is nullified as well, resulting in
\begin{eqnarray}
  &&e^{iE'\hat{z}F_z\delta t'/\hbar}e^{-i\hat{H}'_3\delta t/\hbar}e^{-iE'\hat{z}F_z\delta t'/\hbar}
  \nonumber\\
  &=&\exp\left\{-i\left(\frac{\hbar^2\mathbf{k}^2}{2m}+(\hbar\omega_q+\delta)F_z
  +\hbar\omega_q F_z^2+E_0\right.\right.
  \nonumber\\
  &&\qquad\left.\left.+\frac{\Omega_R}{2}F_x-2\frac{\hbar^22k_L\hat{k}_z}{2m}F_z+E_L\right)
  \delta t/\hbar\right\},
\end{eqnarray}
and its corresponding two-state approximation,
\begin{eqnarray}
  \exp\left\{-i\left(\frac{\hbar^2\mathbf{k}^2}{2m}+
  \frac{\delta}{2}\sigma_z+\frac{\Omega}{2}\sigma_x-2\frac{\hbar^2k_L\hat{k}_z}{2m}\sigma_z\right)
  \delta t/\hbar\right\}. \hskip 12pt
\end{eqnarray}

For the special case of Rashba SOC, the suggested pulse sequence are
illustrated in Fig. \ref{fig2}(b), where the blue and cyan ones are
suitable momentum impulses for compensating the unwanted momentum
recoils in the first and second half cycles respectively. The red
pairs are $\pm\pi/2$ pulses for rotating the pseudo-spin. If the
$\pi/2$ one precedes the $-\pi/2$ pulse, we find in one period
$T=2\delta t$, the total evolution operator under two-state
approximation is given by
\begin{widetext}
\begin{eqnarray}
  U(T,0)&=&e^{i(\sigma_x/2)\pi/2}
  \left(e^{iE'\hat{z}F_z\delta t'/\hbar}e^{-i\hat{H}'_3\delta t/\hbar}e^{-iE'\hat{z}F_z\delta t'/\hbar}\right)
  e^{-i(\sigma_x/2)\pi/2}
  \left(e^{-iE'\hat{y}F_z\delta t'/\hbar}e^{-i\hat{H}_3\delta t/\hbar}e^{iE'\hat{y}F_z\delta t'/\hbar}\right)
  \nonumber\\
  &\simeq&  \exp\left\{-i\left(\frac{\hbar^2\mathbf{k}^2}{2m}+
  \frac{\delta}{2}\sigma_y+\frac{\Omega}{2}\sigma_x-2\frac{\hbar^2k_L\hat{k}_z}{2m}\sigma_y\right)\delta t/\hbar\right\}
  \exp\left\{-i\left(\frac{\hbar^2\mathbf{k}^2}{2m}+\frac{\delta}{2}\sigma_z
   +\frac{\Omega}{2}\sigma_x+2\frac{\hbar^2k_L\hat{k}_y}{2m}\sigma_z\right)\delta t/\hbar\right\}
   \nonumber\\
  &\simeq&  \exp\left\{-i\left(\frac{\hbar^2\mathbf{k}^2}{2m}+
  \frac{\delta}{4}(\sigma_y+\sigma_z)+\frac{\Omega}{2}\sigma_x+\frac{\hbar^2k_L}{2m}(\hat{k}_y\sigma_z-\hat{k}_z\sigma_y)
  \right)2\delta t/\hbar\right\}.
  \label{evolutionop}
\end{eqnarray}
\end{widetext}
According to the Floquet theorem, the quasienergy $\epsilon$ of time-periodic system
is derived from $\det[U(T,0)-e^{-i\epsilon T}]=0$. Then from Eq. (\ref{evolutionop}),
we can easily infer that under first order of $T$ approximation, the quasienergy of our system
is the same as the spectra of that with Rashba SOC.
Reversing the two red $\pm \pi/2$ pulses introduces a minus sign "$-$",
the Rashba SOC then changes into Dresselhaus SOC.
By adjusting the timing constant $\delta t$,
we can extend the above discussion
to SOC of arbitrary form $\beta(\hat{k}_y\sigma_z-\hat{k}_z\sigma_y)+
\sqrt{1-|\beta|^2}\ (\hat{k}_y\sigma_z+\hat{k}_z\sigma_y)$.
The steady state of the effective system Hamiltonian is
reached due to elastic atomic collisions. Although
in the simplest case, one period of the control protocol is often
sufficient, the actual implementation can aim at a higher precision
of the effective SOC Hamiltonian by increasing the number of cycles,
or simple reducing $\delta t$.

A magnetic field gradient was first used in Ref. \cite{lin2009} for
implementing Abelian gauge fields with neutral atoms. However, since
real static B-field is subjected to the Maxwell's equations
$\nabla\cdot\vec B =0$ and $\nabla\times\vec B =0$, one cannot
simply obtain a linear gradient along one direction, e.g. a B-field
like $\vec B=(B_0-by)\hat e_y$ is illegitimate because of its
non-vanishing divergence. The simplest linear gradient B-field,
therefore needs to have two components, like that of the commonly used
two dimensional quadruple field, $\vec B=bx\hat e_x-by\hat e_y$.
When the system is of a reduced dimension, not including
$x$-direction as in Ref. \cite{lin2009}, one is then equipped with a
one-dimensional B-field gradient $\vec B=-by\hat e_y$, which is
equivalent to $\vec B=-by\hat e_z$ upon an axis rotation,
A from precisely needed for implementing $U_1$ above.

Likewise, the above B-field gradient can be simulated using
ac stark shifts from position dependent laser fields far off resonant
coupled to the two states forming an atomic pseudo-spin. Assuming a one-photon
resonant coupling Rabi frequency $\Omega_L(\vec r)$ and a detuning
$\Delta_L=\Omega_L-\Omega_0$, the ac Stark shift takes the form
$\propto \sigma_z|\Omega_L^2(\vec r)|/(2\Delta_L)\propto \sigma_z I_L(\vec r)/\Delta_L$.
A linear spatial gradient can thus be affected with a
laser intensity gradient, which can be implemented using
many methods, including the use of a gradient neutral density filter.
Stronger gradients arise from interfering several waves
forming a standing wave, e.g., with $I_L(\vec r)\sim \cos^2(q y/2)=(1+\cos qy)/2$,
linear gradient is $\propto \pm qy$ around the nodal points of $\cos qy$.

More generally, the state dependent gradients can be engineered to
couple states in the same Zeeman manifold. For example,
the above ac Stark shifts from one-photon coupling can be substituted with
two-photon Raman coupling with suitable differential detuning,
like in Bragg scattering, which then implements impulses $\propto \pm\hat y\sigma_z$,
 $\propto \pm\hat z\sigma_y$ or $\propto \pm\hat z\sigma_z$.

{\it Summarizing} We present a coherent control protocol capable of
realizing the Rashba type SOC in a pseudo-spin 1/2 atomic
quantum gas \cite{lin2011}. For most systems, our protocol can be
implemented in one cycle, involving two separate resonant Raman
coupling. More elaborate forms are possible with multiple control pulses.
When more than one control cycle is implemented, we can
further enhance the precision and strength of the SOC,
or the corresponding artificially created gauge potentials.
In addition, the scheme we suggest is independent of quantum
statistics of atoms, thus can be adopted to fermionic atoms as well.
Our idea thus opens the door for dynamically
implementing artificial gauge potentials in cold atomic systems
based on coherent control theory.

Finally, we compare our idea with two previous schemes
\cite{campbell2011,sau2011}. In Ref. \cite{campbell2011}, three and
four laser fields are needed, cyclically coupled to three or four
internal states. Nearly pure Rashba or
Dresselhaus SOC then results respectively in the limit of large intensity laser
fields. It remains open to find a suitable experimental system.
In Ref. \cite{sau2011},
along each axis of $x$- and $y$- two lasers with different frequency
and their respective reflections are needed. Only in the far-detuned and
small Raman coupling $\Omega_R$ limit, Rashba or Dresselhaus SOC can
be implemented, which results in a relatively small SOC, proportional
to $\Omega_R$. Our scheme, however, takes the full advantage of the
JQI protocol \cite{lin2011}. By simply turning on several pulses,
and making use of the coherent control,
we can dynamically generate arbitrary SOC for neutral atoms.

This work is supported by the NSFC (Contracts No.~91121005 and
No.~11004116). L.Y. is supported by the NKBRSF of China and by the
research program 2010THZO of Tsinghua University.

\end{document}